\documentclass[conference]{IEEEtran}

\usepackage{amsmath}
\usepackage{cite}
\usepackage{graphicx}
\usepackage{epstopdf}
\usepackage{amsfonts,amsmath,amssymb}
\usepackage{graphicx}
\usepackage{url}
\usepackage{bm}
\usepackage{bbm}
\usepackage{subfigure}
\usepackage{stfloats}
\usepackage{citesort}

\newcounter{TempEqysh}

\DeclareMathOperator*{\argmax}{argmax}

\begin{document}

\title{Transmit Antenna Selection with Alamouti Scheme in MIMO Wiretap Channels}

\author{\IEEEauthorblockN{Shihao Yan, Nan Yang, Robert Malaney, and Jinhong Yuan}
\IEEEauthorblockA{School of Electrical Engineering and
Telecommunications, The University of New South Wales, Sydney, NSW,
Australia\\} Email: shihao.yan@student.unsw.edu.au,
nan.yang@unsw.edu.au, r.malaney@unsw.edu.au, j.yuan@unsw.edu.au}

\markboth{Submitted to IEEE GlobeCOM 2013}{Yan
\MakeLowercase{\textit{et al.}}: Transmit Antenna Selection with
Alamouti Scheme in MIMO Wiretap Channels}

\maketitle

\begin{abstract}
This paper proposes a new transmit antenna selection (TAS) scheme which provides enhanced physical layer security in
multiple-input multiple-output (MIMO) wiretap channels. The
practical passive eavesdropping scenario we consider is where
channel state information (CSI) from the eavesdropper is not
available at the transmitter. Our new scheme is carried out in
two steps. First, the transmitter selects the first two strongest
antennas based on the feedback from the receiver, which maximizes
the instantaneous signal-to-noise ratio (SNR) of the
transmitter-receiver channel. Second, the Alamouti scheme is employed at
the selected antennas in order to perform data transmission. At the receiver
and the eavesdropper, maximal-ratio combining is applied in order to exploit
the multiple antennas. We derive a new closed-form expression for the
secrecy outage probability in non-identical Rayleigh fading, and using
this result, we then present the probability of non-zero secrecy capacity
in closed form and the $\varepsilon$-outage secrecy capacity in
numerical form. We demonstrate that our proposed TAS-Alamouti scheme
offers lower secrecy outage probability than a single TAS scheme
when the SNR of the transmitter-receiver channel is above a specific value.
\end{abstract}

\IEEEpeerreviewmaketitle

\section{Introduction}

Physical layer security in wireless communication networks has recently gained considerable research
interest\cite{Shiu}. The core concept behind this paradigm is
to exploit the properties of a wireless channel, such as fading or
noise, to promote secrecy for wireless transmission \cite{Poor}.
 Physical layer security, in principle, eliminates the requirement for complex higher-layer
secrecy techniques, such as encryption and cryptographic key management. In early
pioneering studies
\cite{shannon1949communication,wyner1975the,leung1978gaussian}, the
wiretap channel was characterized as the fundamental framework within which to
protect information at the physical layer. In the wiretap channel
where the transmitter (commonly known as Alice), the receiver (Bob)
and the eavesdropper (Eve) are equipped with a single antenna, it
was proved that perfect secrecy can be achieved if the
eavesdropper's channel between Alice and Eve is worse than the main
channel between Alice and Bob. Motivated by emerging multiple-input
multiple-output (MIMO) techniques, there is a growing interest in
investigating the MIMO wiretap channel
\cite{khisti2010secure,Oggier}. In the MIMO wiretap channel where Alice,
Bob, and/or Eve are equipped with multiple antennas, it was
established that the perfect secrecy can be guaranteed via
beamforming with/without artificial noise even if the quality of the
eavesdropper's channel is higher than that of the main channel
\cite{goel2008guaranteeing,mukherjee2011robust}.




When the CSI of the eavesdropper's channel is
not available at Alice, perfect secrecy can not be
guaranteed. In this circumstance, the secrecy performance is
investigated from the perspective of wireless channel statistics.
Correspondingly, secrecy outage probability is adopted as a
practical and important performance metric to evaluate the
probability that the actual transmission rate is larger than the instantaneous secrecy capacity \cite{bloch2008wireless}. To avoid the high feedback overhead and high signal processing cost required by
beamforming, transmit antenna selection (TAS) was proposed to
enhance physical layer security in MIMO wiretap channels
\cite{alves2012performance,yang2012secure}. In this TAS scheme, only
one antenna is selected at Alice to maximize the instantaneous
signal-to-noise ratio (SNR) of the main channel. Throughout this
paper, we refer to this scheme as \emph{single TAS}. Single TAS
significantly reduces the feedback overhead and hardware complexity,
since only the index of the selected transmit antenna is fed back
from the receiver and only one radio-frequency chain is implemented
at the transmitter. Motivated by this, \cite{alves2012performance}
derived the secrecy outage probability of single TAS for the wiretap
channel with multiple antennas at Alice and Eve but a single antenna
at Bob. A general wiretap channel model with multiple antennas at
Alice, Bob, and Eve was investigated in \cite{yang2012secure}, in
which the performance of single TAS with maximal-ratio combining
(MRC) or selection combining (SC) was thoroughly examined in
Rayleigh fading. Considering versatile Nakagami-$m$ fading,
\cite{yang2012transmit} analyzed the secrecy performance of single
TAS.



In general, the practicality of TAS schemes is determined by not
only the achievable performance but the incurred feedback overhead
and hardware complexity. As such, it is possible to select more than
one antenna at the transmitter for transmission. For such
multi-antenna selection schemes, effective coding strategies need to
be incorporated according to the number of selected antennas
\cite{yuan2006adaptive}. A practical example is the Alamouti scheme
which offers full diversity order \cite{alamouti1998simple}. Against
this background, \cite{chen2003performance} proposed a new
TAS-Alamouti scheme in MIMO systems without secrecy constraints. In
this scheme, two transmit antennas are selected at the transmitter
and the Alamouti scheme is adopted to perform data transmission
through the selected antennas. Notably, it was observed in
\cite{chen2003performance} that the performance of TAS-Alamouti is
worse than that of single TAS in MIMO systems without secrecy
constraints. The reason for this observation lies in the fact that
TAS-Alamouti wastes transmit energy on the second selected antenna.

In this paper, we examine the interesting questions: ``\emph{Is the secrecy performance in MIMO wiretap channels improved if two antennas are selected at the transmitter instead of one? And if so, by how much?}'' In order to address this question,
we propose TAS-Alamouti in MIMO wiretap channels to enhance physical
layer security. In this MIMO wiretap channel, Alice, Bob, and Eve
are equipped with $N_{A}$, $N_{B}$, and $N_{E}$ antennas,
respectively. In our proposed TAS-Alamouti scheme, the first two
strongest transmit antennas are selected at Alice and the Alamouti
scheme is applied at the two selected antennas to carry out data
transmission. At Bob and Eve, MRC is applied to combine the received
signals. To quantify the performance of our scheme, we derive a
new closed-form expression for the secrecy outage probability. Based
on this result, we characterize the probability of non-zero secrecy
capacity and $\varepsilon$-outage secrecy capacity. Our key conclusion
is that our proposed TAS-Alamouti scheme achieves a lower
secrecy outage probability than the single TAS scheme when the SNR
of the main channel is in the medium and high regime relative to the
SNR of the eavesdropper's channel. This useful result is quite a surprising  given the previous studies
of similar schemes in MIMO systems without secrecy constraints \cite{chen2003performance}.


The rest of this paper is organized as follows. Section
\ref{sec_system} details the system model and the proposed
TAS-Alamouti scheme. In Section \ref{sec_performance}, the secrecy
performance of the proposed TAS-Alamouti scheme is analyzed.
Numerical results are presented in Section \ref{sec_numerical}.
Finally, Section \ref{sec_conclusion} draws some concluding remarks and future directions.

\emph{Notation:} Scalar variables are denoted by italic symbols.
Vectors and matrices are denoted by lower-case and upper-case
boldface symbols, respectively. Given a complex vector $\mathbf{x}$,
$\|\mathbf{x}\|$ denotes the Euclidean norm, $(\mathbf{x})^{T}$
denotes the transpose operation, and $(\mathbf{x})^{\dag}$ denotes
the conjugate transpose operation. The $m\times{m}$ identity matrix
is referred to as $\textbf{I}_{m}$.

\section{System Model and Proposed TAS-Alamouti}\label{sec_system}

\begin{figure}[!t]
\begin{center}
{\includegraphics[width=2.2in]{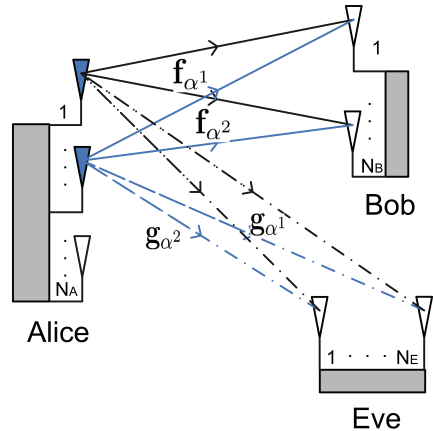}}
\end{center}
\caption{Illustration of a MIMO wiretap channel with $N_A$, $N_B$,
and $N_E$ antennas at Alice, Bob, and Eve, respectively.}
\label{fig:TAS}
\end{figure}

Fig. \ref{fig:TAS} depicts the MIMO wiretap channel of interest,
where the transmitter (Alice), the receiver (Bob), and the
eavesdropper (Eve) are equipped with $N_A$, $N_B$, and $N_E$
antennas, respectively. We assume that the main channel between
Alice and Bob and the eavesdropper's channel between Alice and Eve
are subject to quasi-static Rayleigh fading. Under this assumption,
the fading coefficients are invariant during two adjacent blocks of time
durations within which the Alamouti scheme is applied. We also assume  the same fading block length in the main
channel and the eavesdropper's channel. For such a wiretap channel,
the passive eavesdropping scenario is considered where Eve overhears
the transmission between Alice and Bob without inducing any
interference to the main channel. In this scenario, the
instantaneous channel state information (CSI) of the eavesdropper's
channel is not available at Alice. Of course, we preserve the
assumption that Bob has the full CSI of the main channel and Eve has
the full CSI of the eavesdropper's channel.


We propose a TAS-Alamouti scheme to enhance the physical layer
security in the MIMO wiretap channel of interest. Specifically, the
proposed scheme is performed in two steps. We next detail the two
steps, as follows:

\subsubsection{First Step -- TAS}

In the first step, the first two strongest antennas out of $N_{A}$
antennas are selected at Alice. These two antennas maximize the
instantaneous SNR between Alice and Bob. Here, Bob employs MRC to combine the received signals. As per this criterion, the
index of the first strongest antenna is given by
\begin{equation}\label{AS1}
\alpha_{1}=\argmax_{0\leq\alpha\leq{}N_{A}}\left\|\bm{f}_{\alpha}\right\|
\end{equation}
and the index of the second strongest antenna is given by
\begin{equation}\label{AS2}
\alpha_{2}=\argmax_{0\leq\alpha\leq{}N_{A},\alpha\neq\alpha_{1}}\left\|\bm{f}_{\alpha}\right\|.
\end{equation}
In \eqref{AS1} and \eqref{AS2}, we denote
$\bm{f}_{\alpha}=\left[f_{\alpha,1},f_{\alpha,2},...,f_{\alpha,N_B}\right]^{T}$
as the $N_B \times 1$ channel vector between the $\alpha$-th antenna
at Alice and the $N_B$ antennas at Bob with independent and
identically distributed (i.i.d.) Rayleigh fading entries.

To conduct antenna selection, Alice sends Bob pilot symbols prior to
data transmission. Using these symbols, Bob precisely estimates the
CSI of the main channel and determines $\alpha_{1}$ and $\alpha_{2}$
according to \eqref{AS1} and \eqref{AS2}. After that, Bob feeds back
$\alpha_{1}$ and $\alpha_{2}$ to Alice via a low-rate feedback
channel. As such, our scheme reduces the feedback overhead compared
with beamforming, since only a small number of bits are required to
feedback the antenna indices. We note that the antenna indices
\eqref{AS1} and \eqref{AS2} are entirely dependent on the main
channel. As such, the two strongest transmit antennas for Bob
corresponds to two random transmit antennas for Eve. It follows that
our scheme improves the quality of main channel relative to the
eavesdropper's channel, which in turn promotes the secrecy of the
wiretap channel.


\subsubsection{Second Step -- Alamouti}

In the second step, Alice adopts the Alamouti scheme to perform
secure transmission at the two selected antennas. After receiving
the signals from Alice, Bob applies MRC to combine the received
signals and maximize the SNR of the main channel. This allows Bob to
exploit the $N_{B}$-antenna diversity and maximize the probability
of secure transmission. At the same time, Eve applies MRC to exploit
the $N_{E}$-antenna diversity and maximize the probability of
successful eavesdropping.

As per the rules of the Alamouti scheme, the received signal vectors
at Bob in the first and second time slots are given by
\begin{align}\label{Bob_signal1_Alamouti}
\bm{y}_B(1) = \left[\bm{f}_{\alpha_{1}}, \bm{f}_{\alpha_{2}}\right] \left[\begin{array}{c}x_1 \\
x_2 \end{array}\right] + \bm{n}(1),
\end{align}
and
\begin{align}\label{Bob_signal2_Alamouti}
\bm{y}_B(2) = \left[\bm{f}_{\alpha_{1}}, \bm{f}_{\alpha_{2}}\right] \left[\begin{array}{c}-x_2^{\dag} \\
x_1^{\dag} \end{array}\right] + \bm{n}(2),
\end{align}
respectively, where $\left[\bm{f}_{\alpha_{1}},
\bm{f}_{\alpha_{2}}\right]$ is the $N_B \times 2$ main channel
matrix after TAS, $[x_1, x_2]^T$ is the transmit signal vectors in
the first time slot, $[-x_2^{\dag}, x_1^{\dag}]^T$ is the transmit
signal vectors in the second time slot, and $\bm{n}$ is the
zero-mean circularly symmetric complex Gaussian noise vector
satisfying $\mathbb{E}[\bm{n} \bm{n}^{\dag}] = \bm{I}_{N_B}
\sigma^2$. Under the power constraint, we have
$\mathbb{E}[|x_1|^2]=\mathbb{E}[|x_2|^2]\leq{}P_A/2$, where $P_A$ is
the total transmit power at Alice.

By performing MRC and space-time signal processing, the signals at
Bob are expressed as
\begin{align}\label{Bob_signal1_Alamouti_x}
\bm{y}_B^1 = \left(\bm{f}_{\alpha_{1}}^{\dag}\bm{f}_{\alpha_{1}} +
\bm{f}_{\alpha_{2}}^{\dag}\bm{f}_{\alpha_{2}}\right) x_1 +
\bm{f}_{\alpha_{1}}^{\dag}\bm{n}(1) + \bm{n}(2)^{\dag}
\bm{f}_{\alpha_{2}},
\end{align}
and
\begin{align}\label{Bob_signal2_Alamouti_x}
\bm{y}_B^2 &= \left(\bm{f}_{\alpha_{1}}^{\dag}\bm{f}_{\alpha_{1}} +
\bm{f}_{\alpha_{2}}^{\dag}\bm{f}_{\alpha_{2}}\right) x_2 +
\bm{f}_{\alpha_{2}}^{\dag}\bm{n}(1) - \bm{n}(2)^{\dag}
\bm{f}_{\alpha_{1}}.
\end{align}
Since $\bm{n}(1)$ and $\bm{n}(2)$ are independent from each other,
the instantaneous SNR at Bob for $x_1$ is identical to the
instantaneous SNR at Bob for $x_2$, which is written as
\begin{equation}\label{Bob_SNR}
\gamma_B=\frac{(\|\bm{f}_{\alpha_{1}}\|^2+\|\bm{f}_{\alpha_{2}}\|^2)P_A}{2\sigma^2}.
\end{equation}

Following the same procedure as detailed above, the instantaneous
SNR at Eve is written as
\begin{equation}\label{Eve_SNR}
\gamma_E=\frac{(\|\bm{g}_{\alpha_{1}}\|^2+\|\bm{g}_{\alpha_{2}}\|^2)P_A}{2\sigma^2},
\end{equation}
where $\left[\bm{g}_{\alpha_{1}}, \bm{g}_{\alpha_{2}}\right]$ is the
$N_E \times 2$ eavesdropper's channel matrix after TAS and
$\bm{g}_{\alpha} = [g_{\alpha,1},g_{\alpha,2},...,g_{\alpha,N_E}]^T$
denotes the $N_E \times 1$ channel vector between the $\alpha$-th
antenna at Alice and the $N_E$ antennas at Eve with i.i.d. Rayleigh
fading entries.

\section{Secrecy Performance of TAS-Alamouti}\label{sec_performance}

In this section, we concentrate on the secrecy performance of the
proposed TAS-Alamouti scheme for non-identical Rayleigh fading
between the main channel and the eavesdropper's channel.
Specifically, we derive a new closed-form expression for the secrecy
outage probability. Based on this result, we express the probability
of non-zero secrecy capacity in closed form and present the
$\varepsilon$-outage secrecy capacity in integral form.


\subsection{Secrecy Outage Probability}

The secrecy outage probability is defined as the probability of the secrecy capacity $C_s$ being less than a specific
transmission rate $R_s$ (bits/channel) \cite{bloch2008wireless}. In the MIMO
wiretap channel, $C_{s}$ is expressed as
\begin{equation}\label{secrecy_capacity}
C_{s}=\left\{
\begin{array}{ll}
C_{B}-C_{E}~\;, &  \mbox{$\gamma_{B}>\gamma_{E}$}\\
0~\;, & \mbox{$\gamma_{B}\leq\gamma_{E}$},\;
\end{array}
\right.
\end{equation}
where $C_{B}=\log_{2}\left(1+\gamma_{B}\right)$ is the capacity of
the main channel and $C_{E}=\log_{2}\left(1+\gamma_{E}\right)$ is
the capacity of the eavesdropper's channel. Here, $\gamma_{B}$ is
the instantaneous SNR at Bob given by \eqref{Bob_SNR} and
$\gamma_{E}$ is the instantaneous SNR at Eve given by
\eqref{Eve_SNR}. According to the definition, the secrecy outage
probability is formulated as
\begin{align}\label{outage_probability}
P_{out}\left(R_{s}\right)=\Pr\left(C_{s}<R_{s}\right).
\end{align}

We commence our analysis by presenting the probability density
functions (pdfs) of $\gamma_B$ and $\gamma_E$. Specifically, we
adopt \cite[Eq. (15)]{chen2003performance} as the pdf of $\gamma_B$,
$f_{\gamma_{B}}\left(\gamma\right)$, and define
$\overline{\gamma}_B$ as the average per-antenna SNR at Bob. We note
that $\alpha_{1}$ and $\alpha_{2}$ are equivalent to two random
transmit antennas for Eve. As such, the pdf of $\gamma_E$,
$f_{\gamma_{E}}\left(\gamma\right)$, is given by \cite[Eq.
(15)]{chen2003performance} with $N_{A}=2$. We further define
$\overline{\gamma}_E$ as the average per-antenna SNR at Eve.

We now proceed with the calculation of the secrecy outage
probability. Specifically, we rewrite $P_{out}\left(R_{s}\right)$ as
\begin{align}
P_{out}(R_s) =& \underbrace{P_r(C_s < R_s|\gamma_B > \gamma_E) \times P_r (\gamma_B > \gamma_E)}_{V_1}\notag\\
&+ \underbrace{P_r (\gamma_B < \gamma_E)}_{V_2},
\end{align}
where $V_1$ is derived as
\begin{align}\label{V_1}
V_1 =& \int_0^{\infty}\int_{\gamma_E}^{2^{R_s}(1+\gamma_E) -1}f_{\gamma_E}(\gamma_E)f_{\gamma_B}(\gamma_B)d\gamma_B d\gamma_E\notag\\
=& \underbrace{\int_0^{\infty}\int_{0}^{2^{R_s}(1+\gamma_E) -1}f_{\gamma_E}(\gamma_E)f_{\gamma_B}(\gamma_B)d\gamma_B d\gamma_E}_{U_1}\notag\\
&-\underbrace{\int_0^{\infty}\int_0^{\gamma_E}f_{\gamma_E}(\gamma_E)f_{\gamma_B}(\gamma_B)d\gamma_B
d\gamma_E}_{U_2}.
\end{align}
It is easy to observe that $U_2 = V_2$. As such, we simplify
$P_{out}\left(R_{s}\right)$ as $P_{out}\left(R_{s}\right)=U_{1}$.

\setcounter{TempEqysh}{\value{equation}} \setcounter{equation}{12}
\begin{subequations}
\begin{figure*}[ht]
\begin{small}
\begin{align}\label{L1}
\hspace{0 pt} \Psi_1 &= \sum_{i=0}^{N_A - 2} \sum_{j=0}^{N_B -
1}\sum_{m=0}^{N_E - 1} (-1)^{i+1} \mathbb{G} \sum_{t=0}^{(N_B - 1)i}
a_t(N_B, i) \sum_{k = 0}^{\omega_1 +
k}\left(\frac{2^{R_s}\bar{\gamma}_E}{\bar{\gamma}_B}\right)^{\omega_1}\notag
\frac{k!
\binom{\omega_1 + k}{k}}{2^{\omega_1}(i+2)^{k+1}}e^{-\frac{\left(2^{R_s}-1\right)(i+2)}{\bar{\gamma}_B}}\sum_{u=0}^{\omega_1}\binom{\omega_1}{u}\left(\frac{2 - 2^{1-R_s}}{\bar{\gamma}_E}\right)^{\omega_1 - u}\notag\\
&~~~~~\times\left[ \sum_{n = 0}^{2N_E - m -2} \frac{ n! \binom{2N_E
- m -2}{n}}{2^{2N_E - m-2}}\mathbf{F}_1(\varphi_1) + \sum_{q =
0}^{N_E - m -1}\frac{ (-1)^q \binom{N_E - m -1}{q}}{2^{N_E + q  -
1}(N_E + q)} \mathbf{F}_2(\varphi_1)\right],
\end{align}
\hrule
\end{small}
\end{figure*}


\begin{figure*}[ht]
\begin{small}
\begin{align}\label{L2}
\hspace{0.2 cm}
\Psi_2 &= \sum_{j=0}^{N_B - 1}\sum_{m=0}^{N_E - 1}\mathbb{H}\sum_{p=0}^{N_B \!-\! j \!-\!1}\frac{(-1)^p  \binom{N_B\!-\!j\!-\!1}{p}}{2^{N_B + p -1}(N_B + p)}\left(\frac{2^{R_s}\bar{\gamma}_E}{\bar{\gamma}_B}\right)^{2N_B -j -1}e^{-\frac{\left(2^{R_s + 1}-2\right)}{\bar{\gamma}_B}}\sum_{u=0}^{2N_B -j-1}\binom{2N_B -j-1}{u}\left(\frac{2 - 2^{1-R_s}}{\bar{\gamma}_E}\right)^{2N_B -j-u-1}\notag \\
&~~~~~\times \left[\sum_{n = 0}^{2N_E - m -2} \frac{n! \binom{2N_E -
m -2}{n}}{2^{2N_E - m -1}}\mathbf{F}_1(\varphi_2) +\sum_{q = 0}^{N_E
- m -1}\frac{(-1)^{q}  \binom{N_E -m-1}{q}}{2^{N_E + q}(N_E +
q)}\mathbf{F}_2(\varphi_2)\right],
\end{align}
\hrule
\end{small}
\end{figure*}
%
%
\begin{figure*}[ht]
\begin{small}
\begin{align}\label{L3}
\hspace{-50 cm} \Psi_3 &= \sum_{i=1}^{N_A - 2} \sum_{j=0}^{N_B -
1}\sum_{m=0}^{N_E - 1} (-1)^{i}\mathbb{G} \sum_{t=0}^{(N_B - 1)i}
a_t(N_B, i) \sum_{p=0}^{N_B - j -1}(-1)^p \binom{N_B -j - 1}{p}
\sum_{k = 0}^{\omega_2 + k} \frac{k! \binom{\omega_2 +
k}{k}}{2^{\omega_2}i^{k+1}}\left(\frac{2^{R_s}\bar{\gamma}_E}
{\bar{\gamma}_B}\right)^{\omega_1}e^{-\frac{\left(2^{R_s}-1\right)
(i+2)}{\bar{\gamma}_B}}\notag\\
&~~~~~\times\sum_{u=0}^{\omega_1}\binom{\omega_1}{u}\left(\frac{2 -
2^{1-R_s}}{\bar{\gamma}_E}\right)^{\omega_1 - u}\left[ \sum_{n =
0}^{2N_E - m -2} \frac{ n! \binom{2N_E - m -2}{n}}{2^{2N_E -
m-2}}\mathbf{F}_1(\varphi_1) + \sum_{q = 0}^{N_E - m -1}\frac{
(-1)^q \binom{N_E - m -1}{q}}{2^{N_E + q  - 1}(N_E + q)}
\mathbf{F}_2(\varphi_1)\right],
\end{align}
\hrule
\end{small}
\end{figure*}

\begin{figure*}[ht]
\begin{small}
\begin{align}\label{L4}
\hspace{-50 cm}
\Psi_4 &= \sum_{i=1}^{N_A - 2} \sum_{j=0}^{N_B - 1}\sum_{m=0}^{N_E - 1} (-1)^{i+1}\mathbb{G} \sum_{t=0}^{(N_B - 1)i} a_t(N_B, i)\sum_{p=0}^{N_B - j -1}\frac{(-1)^p \binom{N_B -j - 1}{p}(N_B + p +t -1)!}{i^{N_B +p+t}}\left(\frac{2^{R_s}\bar{\gamma}_E}{\bar{\gamma}_B}\right)^{\omega_3}e^{-\frac{\left(2^{R_s + 1}-2\right)}{\bar{\gamma}_B}}\notag\\
&~~~~~\times \sum_{u=0}^{\omega_3}\binom{\omega_3}{u}\left(\frac{2 -
2^{1-R_s}}{\bar{\gamma}_E}\right)^{\omega_3 - u} \left[\sum_{n =
0}^{2N_E - m -2} \frac{n! \binom{2N_E - m -2}{n}}{2^{2N_E - m
-2}}\mathbf{F}_1(\varphi_2)+\sum_{q = 0}^{N_E - m -1}\frac{(-1)^{q}
\binom{N_E -m-1}{q}}{2^{N_E + q -1}(N_E +
q)}\mathbf{F}_2(\varphi_2)\right].
\end{align}
\hrule
\end{small}
\end{figure*}
\end{subequations}
\setcounter{equation}{\value{TempEqysh}}

To derive $U_1$, we first calculate the inner integral with respect
to $\gamma_B$ by substituting $f_{\gamma_{B}}\left(\gamma\right)$
and $f_{\gamma_{E}}\left(\gamma\right)$ into $U_{1}$ and applying
\cite[Eq. (3.351.1)]{gradshteuin2007table}. We then expand the
product of the inner integral and
$f_{\gamma_{E}}\left(\gamma\right)$ by applying \cite[Eq.
(1.111)]{gradshteuin2007table} and solve the resultant integral with
respect to $\gamma_E$ by applying \cite[Eq.
(3.351.3)]{gradshteuin2007table}. By performing some algebraic
manipulations, the secrecy outage probability is derived as
\begin{equation}\label{outage}
P_{out}(R_s) = 1 - \frac{N_A(N_A - 1)\left[\Psi_1 - \Psi_2 + \Psi_3
- \Psi_4\right]}{\left[\left(N_B - 1\right)!\left(N_E -
1\right)!\right]^2},
\end{equation}
where $\Psi_1$, $\Psi_2$, $\Psi_3$ and $\Psi_4$ are presented in
(\ref{L1}), (\ref{L2}), (\ref{L3}) and (\ref{L4}), respectively. In
(\ref{L1}), (\ref{L2}), (\ref{L3}) and (\ref{L4}), we define
variables $\omega_1 = 2N_B + t-j-k-2$, $\omega_2 = N_B + p+t-1$,
$\omega_3 = N_B -j-p-1$, $\lambda = 2N_E +u-m-n-3$, $\varphi_1 =
\frac{\overline{\gamma}_B +
2^{R_s-1}(i+2)\overline{\gamma}_E}{\overline{\gamma}_B}$, and
$\varphi_2 = \frac{\overline{\gamma}_B +
2^{R_s}\overline{\gamma}_E}{\overline{\gamma}_B}$. We also define
the functions
\begin{equation}
\mathbb{G}=\binom{N_A-2}{i}j!\binom{N_B-1}{j}m!\binom{N_E-1}{m},\notag
\end{equation}
\begin{equation}
\mathbb{H}=j!\binom{N_B - 1}{j}m!\binom{N_E - 1}{m},\notag
\end{equation}
\begin{equation}
\mathbf{F}_1(\varphi) = (2N_E -m -n -2)\mathbf{W}\left(\lambda,
\varphi\right)- \mathbf{W}\left(\lambda + 1, \varphi\right),\notag
\end{equation}
and
\begin{equation}
\mathbf{F}_2(\varphi) = (2N_E -m -2)\mathbf{W}\left(\lambda,
\varphi\right)- \mathbf{W}\left(\lambda + 1, \varphi\right).\notag
\end{equation}
We further define $a_t(N_B,i)$ as the coefficients of $z^t$ for $0
\leq t \leq i(N_B -1)$, which arises from the expansion of
\begin{equation}
\left(\sum_{k = 0}^{N_b -1}\frac{z^k}{k!}\right)^i,
\end{equation}
and define $\mathbf{W}(r,u)$ for $u > 0$ in $\mathbf{F}_1(\varphi)$
and $\mathbf{F}_2(\varphi)$ as
\begin{equation}\label{W_function}
\mathbf{W}(r,u) \!=\! \int_0^{\infty} x^r e^{-ux} dx \!=\!
\begin {cases}
\begin{split}
    &r! u^{-r-1},~\textrm{if} ~~r = 0, 1, 2,...\\
    &0, ~~~~~~~~~\textrm{if}~~r = -1.
\end{split}
\end {cases}
\end{equation}
It is highlighted that our new expression in \eqref{outage} is in
closed form as it involves finite summations of exponential
functions and power functions.


\subsection{Probability of Non-zero Secrecy Capacity}

The probability of non-zero secrecy capacity is defined as the
probability by which the secrecy capacity is larger than zero. As
such, it is expressed as
\begin{align}\label{non-zero}
\Pr\left(C_s>0\right)&= \Pr\left(\gamma_B > \gamma_E\right)\notag\\
&= \int_0^{\infty} \int_0^{\gamma_B}
f_{\gamma_B}\left(\gamma_B\right)f_{\gamma_E}\left(\gamma_E\right)d\gamma_E
d\gamma_B.
\end{align}
Substituting $f_{\gamma_{B}}\left(\gamma\right)$ and
$f_{\gamma_{E}}\left(\gamma\right)$ into \eqref{non-zero} and
solving the resultant integrals, the explicit expression for
$\Pr\left(C_s>0\right)$ is obtained. Due to page limits, the
explicit expression is omitted here. Instead, we present
$\Pr\left(C_s>0\right)$ in terms of the secrecy outage probability
as
\begin{equation}\label{non-zero_s}
\Pr\left(C_s>0\right)=1-P_{out}\left(0\right).
\end{equation}

\subsection{$\varepsilon$-Outage Secrecy Capacity}

The $\varepsilon$-outage secrecy capacity is defined as the maximum
secrecy rate for which the secrecy outage probability is no larger
than $\varepsilon$. Specifically, it is characterized as
\begin{equation}\label{outage_capacity}
C_{out}\left(\varepsilon\right)=\argmax_{P_{out}\left(R_{s}\right)\leq\varepsilon}R_{s}.
\end{equation}
Substituting \eqref{outage} into \eqref{outage_capacity} and
applying numerical root finding, $C_{out}\left(\varepsilon\right)$
is obtained.

\section{Numerical Results}\label{sec_numerical}

In this section, we present numerical results to examine the impact
of the number of antennas and the average SNRs on the secrecy
performance. Specifically, we conduct a thorough performance
comparison between our TAS-Alamouti scheme with the single TAS
scheme in \cite{yang2012secure}. This comparison highlights the
potential of TAS-Alamouti.

\begin{figure}[t]
\begin{center}
{\includegraphics[width=3.0in, angle =0]{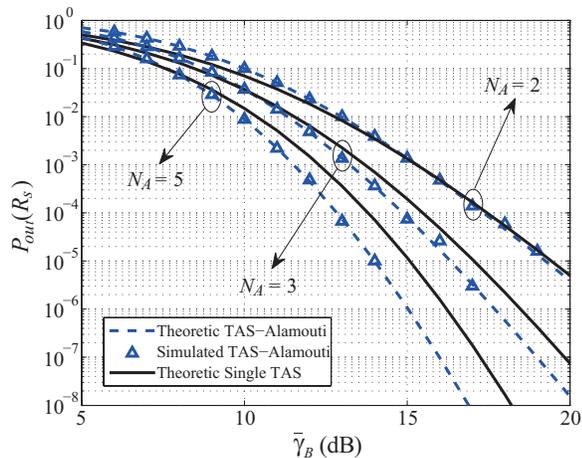}}
\end{center}
\caption{Secrecy outage probability versus $\overline{\gamma}_B$ for
$R_s=1$, $\overline{\gamma}_E=5$ {dB}, $N_B = 3$, and $N_E = 2$.}
\label{fig:outage_NA}
\end{figure}

We first examine the impact of $N_{A}$ on the secrecy outage
probability.  Fig.~\ref{fig:outage_NA} plots
$P_{out}\left(R_{s}\right)$ versus $\overline{\gamma}_B$ for
different $N_{A}$. In this figure, the theoretical TAS-Alamouti curve is
generated from \eqref{outage}, and the theoretical single TAS curve is
generated from \cite[Eq. (13)]{yang2012secure}. We first observe
that $P_{out}\left(R_{s}\right)$ of TAS-Alamouti significantly
decreases as $N_A$ increases. Moreover, we observe that TAS-Alamouti
achieves a lower $P_{out}\left(R_{s}\right)$ than single TAS when
$\overline{\gamma}_B$ is in the medium and high regime. For example,
TAS-Alamouti outperforms single TAS when $\overline{\gamma}_B>10$ dB
for $N_{A}=3$. This is due to the fact that more transmit energy is
wasted on the second strongest antenna for Eve than for Bob at
medium and high $\overline{\gamma}_B$. Furthermore, we observe that
TAS-Alamouti provides a higher $P_{out}\left(R_{s}\right)$ than
single TAS when $\overline{\gamma}_B$ is low. As such, it is easy to
determine the crossover point at which TAS-Alamouti and single TAS
achieve the same performance. Notably, we find that the value of
$\overline{\gamma}_B$ at the crossover point decreases as $N_A$
increases. This can be explained by the fact that the two selected
transmit antennas are determined by Bob and the freedom of
$\alpha$ increases with $N_A$. Finally, we observe that the
theoretical curves match precisely with the Monte Carlo simulations. This verifies the correctness
of our analysis. Monte Carlo simulations are omitted in other
figures to avoid cluttering.

\begin{figure}[t]
\begin{center}
{\includegraphics[width=3.0in, angle =0]{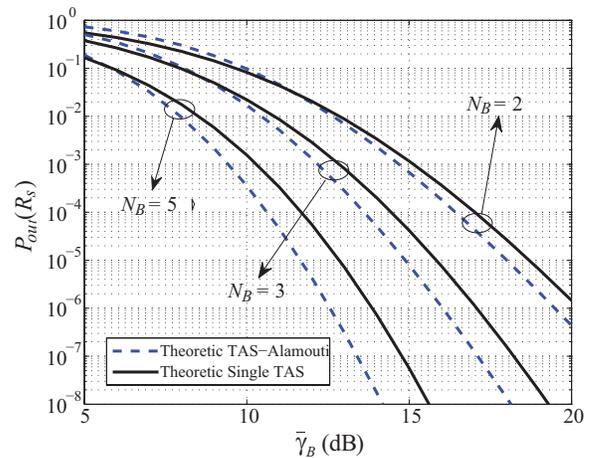}}
\end{center}
\caption{Secrecy outage probability versus $\overline{\gamma}_B$ for
$R_s = 1$, $\overline{\gamma}_E = 5$ {dB}, $N_A = 4$, and $N_E =
2$.} \label{fig:outage_NB}
\end{figure}

\begin{figure}[t]
\begin{center}
{\includegraphics[width=3.0in, angle =0]{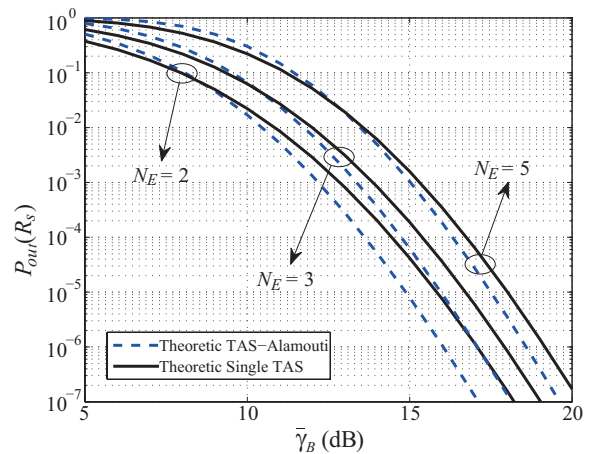}}
\end{center}
\caption{The secrecy outage probability versus $\overline{\gamma}_B$
for $R_s = 1$, $\overline{\gamma}_E =~5$~{dB}, $N_A = 4$, and $N_B
= 3$.} \label{fig:outage_NE}
\end{figure}

We next examine the impact of $N_B$ and $N_E$ on the secrecy outage
probability. Fig.~\ref{fig:outage_NB} plots
$P_{out}\left(R_{s}\right)$ versus $\overline{\gamma}_B$ for
different $N_{B}$. In this figure, we observe that
$P_{out}\left(R_{s}\right)$ of TAS-Alamouti profoundly decreases as
$N_B$ increases. Moreover, we observe that the value of
$\overline{\gamma}_B$ at the crossover point decreases as $N_B$
increases. This is due to the fact that larger $N_B$ increases the freedom
of $\bm{f}_{\alpha}$. Fig.~\ref{fig:outage_NE} plots
$P_{out}\left(R_{s}\right)$ versus $\overline{\gamma}_B$ for
different $N_{E}$. In this figure, we observe that
$P_{out}\left(R_{s}\right)$ of TAS-Alamouti increases as $N_E$
increases. In addition, we observe that the value of
$\overline{\gamma}_B$ at the crossover point increases as $N_E$
increases. This arises from the fact that larger $N_E$ increases the
freedom of $\bm{g}_{\alpha}$.


\begin{figure}[t]
\begin{center}
{\includegraphics[width=3.0in, angle =0]{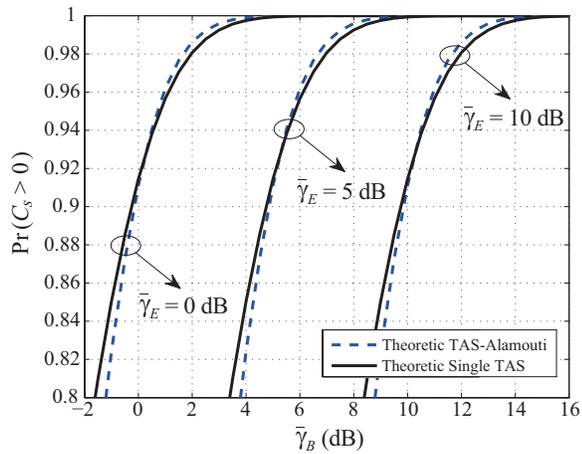}}
\end{center}
\caption{The probability of non-zero secrecy capacity versus $\overline{\gamma}_B$
for $N_A = 4$, $N_B = 3$, and $N_E = 2$.} \label{fig:non}
\end{figure}

We now focus our attention on the probability of non-zero secrecy
capacity. Fig. \ref{fig:non} plots $P_{out}\left(R_{s}\right)$
versus $\overline{\gamma}_{B}$ for different
$\overline{\gamma}_{E}$. In this figure, the theoretical
TAS-Alamouti curve is generated from \eqref{non-zero}, and the theoretical
single TAS curve is generated from \cite[Eq. (29)]{yang2012secure}. From
Fig. \ref{fig:non}, we see that the $\Pr\left(C_{s} > 0\right)$ of
TAS-Alamouti is higher than that of single TAS when
$\overline{\gamma}_{B}$ is larger than a specific value. In
particular, we observe that the value of $\overline{\gamma}_B$ at
the crossover point is around $\overline{\gamma}_E$. Notably, this
value increases as $\overline{\gamma}_E$ increases. Additionally, we
observe that $\Pr\left(C_{s} > 0\right)$ of TAS-Alamouti still
exists when $\overline{\gamma}_B < \overline{\gamma}_E$.

\begin{figure}[t]
\begin{center}
{\includegraphics[width=3.0in, angle =0]{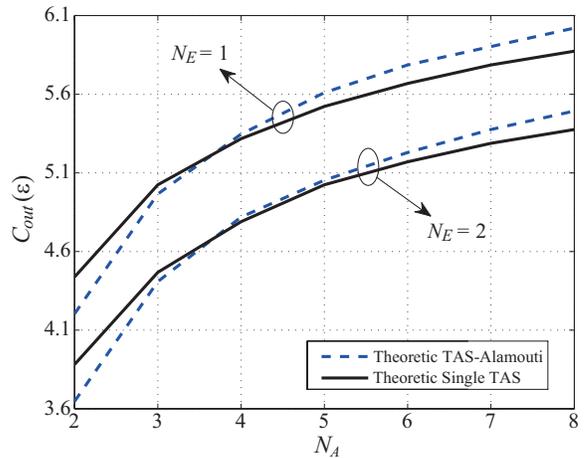}}
\end{center}
\caption{The $\varepsilon$-outage secrecy capacity versus $N_A$ for
$\varepsilon = 0.01$, $\overline{\gamma}_B = 20$~{dB},
$\overline{\gamma}_E = 0$~{dB}, and $N_B = 2$.}
\label{fig:outage_capacity}
\end{figure}

Finally, we examine the $\varepsilon$-outage secrecy capacity. Fig.
\ref{fig:outage_capacity} plots $C_{out}(\varepsilon)$ versus $N_A$
for different $N_{E}$. In this figure, the theoretical TAS-Alamouti curve
is generated from \eqref{outage_capacity}, and the theoretical
single TAS curve is generated from \cite[Eq. (31)]{yang2012secure}. From
this figure, we see that $C_{out}(\varepsilon)$ of TAS-Alamouti
increases with $N_A$ but decreases with $N_E$. We also see
 that TAS-Alamouti achieves a higher $C_{out}(\varepsilon)$
than single TAS when $N_A$ is larger than a certain value.

\section{Conclusion}\label{sec_conclusion}

In this work we have introduced a new  TAS-Alamouti scheme for physical
layer security enhancement in MIMO wiretap channels. Adopting
non-identical Rayleigh fading between the main channel and the
eavesdropper's channel, we derived a new closed-form expression for
the secrecy outage probability, based on which the probability of
non-zero secrecy capacity and $\varepsilon$-outage secrecy capacity
were characterized. We proved that the our TAS-Alamouti scheme
achieves lower secrecy outage probability than the single TAS scheme
when the SNR of the main channel is in the medium and high regime
relative to the SNR of the eavesdropper's channel. Future directions for our new scheme include its integration with location verification techniques \cite{malaney2007securing}\cite{yan2012information} for even more enhanced security at the wireless physical layer.



\section*{Acknowledgments}

This work was funded by The University of New South Wales and
Australian Research Council Grant DP120102607.

\end{document}